\documentclass[
  ,draft            
  ]
  {aipproc}
\input{babarsym.tex}
\layoutstyle{8x11double}

\begin{document}

\title{Recent Results from D0}

\author{Vivek Jain 
\\ {\normalsize\sl (Representing the D0 Collaboration)}
}{
address={Brookhaven National Laboratory, Department of Physics, 
         Upton, NY 11973, U.S.A.}
}

\begin{abstract}

We discuss recent ($B$ physics) results from D0. The results presented
here correspond to an integrated luminosity of $\approx 115 $
pb$^{-1}$ of data collected at the Tevatron, between April  2002 and
June 2003, at a center of mass energy of (\proton \antiproton
collisions) of 1.96 TeV.

\end{abstract}

\maketitle


\section{Introduction}

The $B$ physics program at D0, is designed to be complementary to the
program at the $B$-factories at SLAC and KEK and includes studies of \Bs
oscillations, quarkonia (\jpsi, $\chi_c$, $\Upsilon$), searches for rare
decays such as $B_s \rightarrow \mu^+ \mu^-$, $B$ spectroscopy,
{\it e.g.,} $B^*_{sJ}$, lifetimes of the different $B$ hadrons,
search for the lifetime difference in the \Bs CP eigenstates, study of
beauty baryons, $B_c$, b production cross-section, etc.

One of the more important topics in $B$ physics is the search for
\Bs\Bsb mixing.  Global fits to the unitarity triangle, assuming that the
Standard Model is correct, indicate that the 95\% CL interval for the
mixing parameter $\Delta M_s$ is (14.2-28.1)ps$^{-1}$
\cite{CERN:yellow}.  
The current limit\footnote{The limit is derived by combining limits
from 13 different measurements} is $\Delta M_s > 14.9 {\rm ps}^{-1}$
at the 95\% CL \cite{CERN:yellow}. A measured value of $\Delta M_s$ 
much larger than the upper limits given here could pose a
problem for the Standard Model.

\section{D0 Detector}

For the current run of the Tevatron (Run II), the D0 detector went
through a major upgrade. The inner tracking system was completely
revamped. The detector now includes a Silicon tracker, surrounded by
Scintillating Fiber tracker, both of which are enclosed in a 2 Tesla
solenoidal magnetic field. Pre-shower counters are located before the
calorimeter to aid in electron and photon identification. The muon
system has been improved, {\it e.g.} more shielding to reduce beam
background was added.

The D0 detector has excellent tracking and lepton acceptance. Tracks
with pseudo-rapidity ($\eta$) as large as 2.5--3.0 ($\theta \approx
10^{\circ}$) and transverse momentum (p$_T$) as low as 180 MeV/c are
reconstructed.

The muon system can identify muons within $|\eta| < 2.0$. The minimum
p$_T$ of the reconstructed muons varies as a function of $\eta$. In
most of the results presented here, we required muons to have p$_T >
2$ GeV/c.

Low pT electron identification is currently limited to p$_T > 2$ GeV/c
and $|\eta| < 1.1$; however, we are working to improve both the
momentum and $\eta$ coverage.

A Silicon based (hardware) trigger is being commissioned which will
allow us to trigger on long-lived particles, such as the daughters of
charm and beauty hadrons. We expect to start including this trigger in
the online system after the current Tevatron shutdown ends in
Mid-November. We are currently making impact parameter cuts at Level 3
(software trigger). The hardware trigger will allow us to make these
cuts at an earlier stage.
  
\section{Data Sample}

The results presented here are based on data collected between April
2002 and June 2003. The dataset correspond to an integrated luminosity
of about 115 pb$^{-1}$. We collected data with a dimuon as well as a
single muon trigger. To reduce the data rate, a luminosity dependent
prescale was applied to the single muon trigger (the prescale was 1
for instantaneous ${\cal L} <20\times10^{30} {\rm cm}^{-2} {\rm
s}^{-1}$). In both these triggers, the requirement that muons have
hits in all layers of the muon system implies that they have total
momentum $ \ge 3$ GeV.

\section{Results}

\Bs oscillations is one of three processes where the flavour of the 
initial state changes by two units, the others being $K{\bar K}$ and
$B^0_d{\bar B^0_d}$ mixing. Studies of $K{\bar K}$ mixing yielded
results on indirect and direct CP violation, whereas the (unexpected)
large value of $B^0_d{\bar B^0_d}$ mixing implied that the top quark
was much heavier than expectations. If our current understanding of
the Standard Model and the CKM matrix are correct, then \Bs
oscillations should be in the allowed region, and any deviation could
be a sign of new physics.  If the mixing parameter, $\Delta M_s$ is
very large then the difference in the widths of the CP eigenstates of
the \Bs may be detectable.

The significance of a $B$ mixing measurement can be expressed as,
\begin{equation}
{\rm Sig.} = \sqrt{\frac{N\epsilon D^2}{2}}\exp^{-(\Delta M_s\times \sigma_t)^2/2}\sqrt{\frac{S}{S+B}}
\end{equation}

where $N$ is the number of reconstructed \Bs events, $\epsilon D^2$ is
a measure of how well we know the flavour of the \Bs at production,
$\sigma_t$ is the proper time resolution, and $S/(S+B)$ expresses the
cleanliness of the signal.

To study \Bs oscillations we therefore need three ingredients, (a)
final state reconstruction, (b) ability to measure $B$ decay lengths,
and (c) flavour tagging of the $B$ both at production and decay\footnote{The
latest results can be found at \url{http://www-d0.fnal.gov/Run2Physics/ckm/}}.

\subsection{Final State Reconstruction}

We can reconstruct \Bs in both the hadronic and semi-leptonic final
states. The advantage of the former is that since it is a fully
reconstructed decay, the proper time resolution is much better,
whereas the branching fraction for the latter is much larger ($\approx
10\%$ vs. 0.5\%).

\begin{figure}
  \includegraphics[height=.3\textheight]{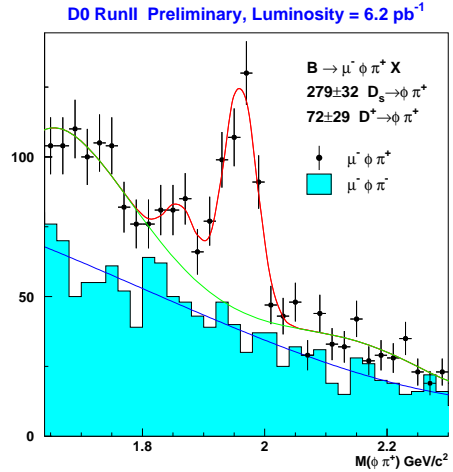}
  \caption{\Bs semi-leptonic yield}\label{BsSL}
\end{figure}

In Fig.~\ref{BsSL} we show the $B_s \rightarrow D_s
\mu X$ signal for ${\int {\cal L}} \approx 6.2$ pb$^{-1}$. 
The blue shaded histogram represents the wrong sign combinations.
By comparison, we expect to see about 1.0 
$B_s \rightarrow D_s \pi$ events/pb$^{-1}$. 

\begin{figure}
  \includegraphics[height=.3\textheight]{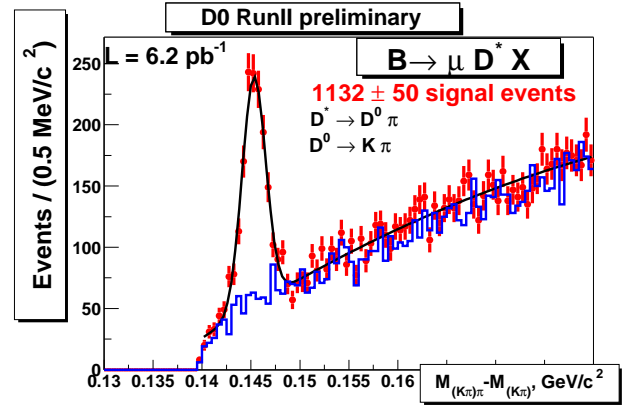}
  \caption{$B\rightarrow D^{*-} \mu X$ }\label{BdSL}
\end{figure}

In Figure~\ref{BdSL}, we present the inclusive $B \rightarrow
D^{*-} \mu X$ signal for ${\int {\cal L}} \approx 6.2$
pb$^{-1}$. The blue line represents the wrong sign combinations. We
can use this mode to study $B^0_d$ mixing and for measuring
inclusive $B$ lifetime.

We will also reconstruct hadronic \Bs decays, {\it e.g.,} $B_s
\rightarrow D_s \pi$, as well as semi-electronic final states.
Details about $B$ hadron reconstruction at D0 can be found in E. Von Toerne's 
contribution to these proceedings.

\subsection{Lifetime Measurement}

\begin{figure}
  \includegraphics[height=.3\textheight]{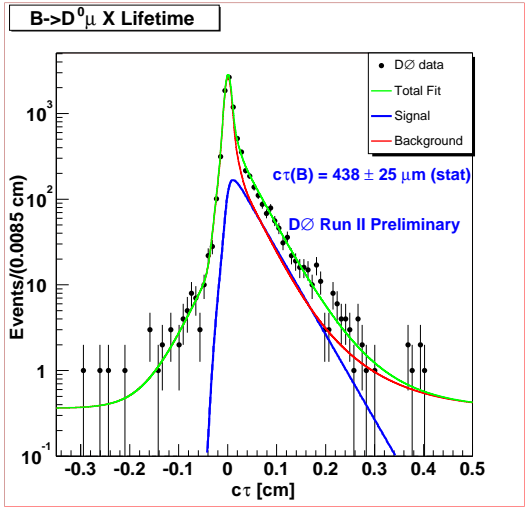}
  \caption{Lifetime for $B\rightarrow D^0 \mu X$ }\label{BSL_ctau}
\end{figure}

In Fig.~\ref{BSL_ctau}, we present a measurement of the inclusive $B$
lifetime using $B\rightarrow D^0 \mu X$ decays.

The background shape is obtained from the \Dz mass sidebands. Both
sidebands and the signal regions are fitted with a combination of a
Gaussian function to represent zero-lifetime background and
exponentials (convoluted with Gaussian resolutions functions) to
represent non-zero lifetime (the convolution is to take account of the
smearing).

From the fit, we measure the inclusive $B$ lifetime to be $438 \pm 25
({\rm stat.})$ $\mu {\rm m}$, which agrees with the world average,
$472 \pm 2$ $\mu {\rm m}$ \cite{PDG:2002}. We have also measured
lifetimes for other $B$ hadrons, and details can be found in Daria
Zieminska's contribution to these proceddings.

\subsection{Flavour Tagging}

To be able to do a mixing measurement, one needs to know the flavour
of the $B$ hadron at the time of production and decay. By using
flavour-specific decays, one can easily tag the flavour at the time of
decay. Tagging the $B$ flavour at production is more work. At D0, we
plan to use the following techniques,
\begin{itemize}
\item
   Soft Lepton Tagging: The sign of the lepton produced in the
   semi-leptonic decay of the other $B$ in the event is used to tag
   the flavour of the other $B$. We then make the assumption that the
   flavour of the decay $B$ is opposite to that of the tag $B$. There
   will be some contamination due to the fact that the other $B$ can
   mix, or that we pick up the tag lepton from a charm semi-leptonic
   decay. This method has low efficiency, but very high tagging
   power. We also plan to use electrons.

\item
   Jet Charge Tagging: We take all tracks opposite to the decay $B$
   and form a track jet.  The assumption is that these tracks are
   produced in the fragmentation of the other b-quark, as well as in
   the decay of the other $B$ hadron. This method has high efficiency,
   but has poorer tagging power.

\item 
   Same Side Tagging: In this technique, we identify tracks produced
   in the fragmentation of the b-quark which gives rise to the decay
   $B$. In addition, the decay B can come from a resonance, {\it
   e.g.,} B$^{**+} \rightarrow B^0\pi^+$, and one can use such pions
   to identify the flavour of the decay B at the time of
   production. This method has high efficiency, but has poorer tagging
   power

\end{itemize} 

We benchmarked these techniques using a sample of $B^+
\rightarrow {\rm J}/\psi {\rm K}^+$ decays. Since the $B^+$ does not
oscillate, it provides for a good testing ground.

\begin{figure}
  \includegraphics[height=.3\textheight]{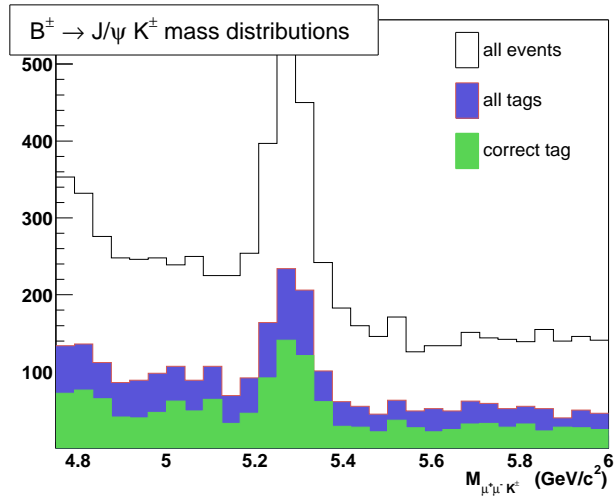}
  \caption{Fits for Jet Charge tagging results}\label{JetQ}
\end{figure}

In Figure~\ref{JetQ}, we show the $B^+$ peak for all events, events
with a (jet-charge) tag, and events with the correct tag. We use the fits to these
peaks to determine the efficiency ($\epsilon$) and tagging power
(Dilution) of the various techniques. $\epsilon = (N_R + N_W)/N_{all}$,
and Dilution = $(N_R - N_W)/(N_R + N_W)$, where $N_{all}$, $N_R$ and
$N_W$ are the number of all events, correctly tagged and wrongly
tagged events, respectively.

The results for the various techniques are summarized in
Table~\ref{tag_results}. We are in the process of checking these
results on a sample of \Bs and
\Bz events.

\begin{table}
\begin{tabular}{lccc}
\hline
  & \tablehead{1}{r}{b}{$\epsilon$}
  & \tablehead{1}{r}{b}{Dilution(D)}
  & \tablehead{1}{r}{b}{$\epsilon D^2$} \\
\hline
Soft Muon & 5\% & 57\% & $1.6\pm 1.1$\% \\
Jet Charge & 47\% & 27\% &$3.3\pm 1.7$\% \\
Same Side & 79\% & 26\% & $5.5\pm 2.0$\% \\
\hline
\end{tabular}
\caption{Summary of flavour tagging results}
\label{tag_results}
\end{table}

\subsection{\Bs mixing}

We expect to observe the following four classes of events,
\begin{itemize}
\item
   $B_s \rightarrow D_s \mu X$ collected with the
   single muon trigger. The flavour at production will be tagged using
   all three techniques discussed above.
\item
   $B_s \rightarrow D_s \pi$ collected with the single
   muon trigger. In this case, we can use the trigger muon as the
   flavour tag. The tagging power will be very high in this case.
\item
   $B_s \rightarrow D_s \mu X$ collected with the
   dimuon trigger. The flavour at production will be tagged using
   the second muon in the event.
\item
   $B_s \rightarrow D_s e X$ collected with the
   single muon trigger. The flavour at production will be tagged using
   the trigger muon.
\end{itemize}

The semi-leptonic events have poorer proper time resolution compared
to the hadronic mode, but much larger statistics. We are in the
process of improving our estimate of the proper time resolution for
both hadronic and semi-leptonic events. In addition, we can combine
all modes to get a better measurement or a better limit.

If, we assume a resolution of $\approx 150$ fs, we project that using
the first class of events (with $\int {\cal L} \approx 500$ pb$^{-1}$), we
can make a 3$\sigma$ measurement if $\Delta M_s
\approx 12$ ps$^{-1}$ or a 1.5$\sigma$ measurement if $\Delta M_s
\approx 15$ ps$^{-1}$. The third and fourth class of events have a
very similar reach in $\Delta M_s$.

For hadronic events (with the same luminosity), if the resolution is
$\approx 110$ fs, the corresponding numbers are a 3$\sigma$
measurement if $\Delta M_s \approx 12$ ps$^{-1}$ or a 2.2$\sigma$
measurement if $\Delta M_s \approx 15$ ps$^{-1}$.

\subsection{Quarkonia}

At the Tevatron, J/$\psi$ can be produced either promptly (direct as
well as indirect), i.e., $\proton \antiproton \rightarrow {\rm J}/\psi
({\rm or } \chi_c \rightarrow {\rm J}/\psi)$ or as a product in $B$
decays. $\Upsilon$ by contrast are only produced promptly.  The study
of quarkonia sheds light on the strong interaction, especially
non-perturbative QCD\footnote{ An excellent review, covering both
theory and experiment, can be found at
\url{http://hep.physics.indiana.edu/~zieminsk/talks.html}. Select the
Trento workshop talk}

In Run I (at the Tevatron), the production cross-section of direct
J/$\psi$ was about 50 times larger than the Color Singlet
Model.

With the new dataset, we plan to update the results on J/$\psi$
production cross-section as a function of $p_T$ and $\eta$, as well as
study polarization effects.  In addition, we are working to measure
the absolute cross-section and polarization of the $\Upsilon$ states.

\section{Conclusions}

The D0 detector has started to produce results in the field of $B$
physics. We currently have $\approx 220$ pb$^{-1}$ of data on tape,
and expect to get up to $\approx 500$ pb$^{-1}$ by the end of 2004.

As a stepping stone to \Bs mixing, we plan to measure $\Delta M_d$ to
benchmark our analyses techniques. In addition, we are pursuing a
vigorous program, which includes measurement of $B$ lifetimes, rare
decays, studies of quarkonia, beauty baryons, $B_c$, etc.


\begin{theacknowledgments}

I would like to acknowledge members of the B group at D0 for
assisting me with this talk, especially Brad Abbott, Christos
Leonidopoulos and Rick Van Kooten.

I would also like to thank the organizers for a very well organized
and stimulating conference.

This work was supported by the U.S. Department of Energy under
Contract No. DE-AC02-98CH10886.

\end{theacknowledgments}


\bibliographystyle{aipproc}   

\bibliography{Beauty2003_vivek}

\hyphenation{Post-Script Sprin-ger}
\begin{thebibliography}{2}
\expandafter\ifx\csname natexlab\endcsname\relax\def\natexlab#1{#1}\fi
\providecommand{\enquote}[1]{``#1''}
\expandafter\ifx\csname url\endcsname\relax
  \def\url#1{\texttt{#1}}\fi
\expandafter\ifx\csname urlprefix\endcsname\relax\def\urlprefix{URL }\fi

\bibitem[M.~Battaglia(2003)]{CERN:yellow}
M.~Battaglia, {\it etal}., {The CKM matrix and the unitarity triangle}, Tech.
  rep., arXiv:hep-ph/0304132 (2003).

\bibitem[K.~Hagiwara(2002)]{PDG:2002}
K.~Hagiwara, {\it etal}., \emph{Physical Review D}, \textbf{66}, 1 (2002).

\end{thebibliography}

\IfFileExists{\jobname.bbl}{}
 {\typeout{}
  \typeout{******************************************}
  \typeout{** Please run "bibtex \jobname" to optain}
  \typeout{** the bibliography and then re-run LaTeX}
  \typeout{** twice to fix the references!}
  \typeout{******************************************}
  \typeout{}
 }

\end{document}